\documentclass[manuscript, nonacm, timestamp, review=false]{acmart}
\AtBeginDocument{%
  \providecommand\BibTeX{{%
    \normalfont B\kern-0.5em{\scshape i\kern-0.25em b}\kern-0.8em\TeX}}}

\setcopyright{acmcopyright}
\copyrightyear{2022}
\acmYear{2022}
\acmDOI{XXXXXXX.XXXXXXX}

\usepackage{xcolor}
\usepackage{multirow}
\usepackage{makecell}
\usepackage{nicematrix}
\usepackage[hypcap=false]{caption}
\usepackage{subcaption}

\acmBooktitle{In review - authors copy}
\acmPrice{15.00}
\acmISBN{978-1-4503-XXXX-X/18/06}

\begin{document}

\title{Aligning Learners' Expectations and Performance by Learning Analytics System with a Predictive Model}


\author{Saša Brdnik}
\affiliation{%
  \institution{University of Maribor}
  \city{Maribor}
  \country{Slovenia}}
\email{sasa.brdnik@um.si}

\author{Bostjan Šumak}
\affiliation{%
  \institution{University of Maribor}
  \city{Maribor}
  \country{Slovenia}
}
\email{bostjan.sumak@um.si}

\author{Vili Podgorelec}
\affiliation{%
  \institution{University of Maribor}
  \city{Maribor}
  \country{Slovenia}
}
\email{vili.podgorelec@um.si}


%

\begin{abstract}
Learning analytics (LA) is data collection, analysis, and representation of data about learners in order to improve their learning and performance. Furthermore, LA opens the door to opportunities for self-regulated learning in higher education, a circular process in which learners activate and sustain behaviors that are systematically oriented toward their personal learning goals. The potentials of LA and self-regulated learning are huge; however, they are not yet widely applied in higher education institutions.  Slovenian higher education institutions have lagged behind other European countries in LA adoption. Our research aims to fill this gap by using a qualitatively and quantitatively led workflow for building requirement-oriented LA solution, consisting of empirically gathering the students' expectations of LA and presenting a dashboard solution. Translated Student Expectations of Learning Analytics Questionnaire (SELAQ) and focus groups were used to gather expectations from learners. Based on this data, a user interface utilizing learning analytics and grade prediction with an artificial intelligence model was implemented for a selected course. The interface includes early grade prediction, peer comparison, and historical data overview. Early grade prediction is based on a machine learning model built on users' interaction in the virtual learning environment, demographic data and their lab grades. First, classification is used to determine students at risk of failing - its precision is reaching 98\% after the first month of the course. Second, the exact grade is predicted with the Decision Tree Regressor, which reaches a mean absolute error of 11.2grade points (on a 100 points scale) after the first month. The proposed system is designed for students - its main benefit is the support for self-regulation of the learning process during the semester, possibly motivating students to adjust their learning strategies to prevent failing the course. Initial student evaluation of the system showed positive results.



\end{abstract}

\keywords{learning analytics, student expectations, self-regulated learning, explainable artificial intelligence, academic success prediction}

\received{14 October 2023}

\maketitle

\section{Introduction}


During the pandemic, the majority of higher education institutions increased the integration of virtual learning environments in their learning process. The possibilities of using the abundance of gathered data to improve the learning process for students have immensely increased. The main stakeholders in learning analytics (LA) solutions are teachers and learners; this work is focused on expectations and solutions for the latter. This study builds on previous research about students' expectations and attitudes in Europe \cite{Engström2022}\cite{SELAQ-multilang}\cite{Nouri2019} and contributes to the recognized importance of including students as key stakeholders in the design and implementation of LA \cite{Tsai2020}. We focus on the context of Slovenian higher engineering education, which has not been explored in this context before. We contribute to the current body of research with a LA solution based on student requirements, gathered with quantitative and qualitative methods. Such an approach has seldom been combined in existing work; studies mostly gather stakeholders' expectations \cite{SHEILA2017}\cite{Garcia2021}\cite{SELAQ-multilang} or provide finalized LA solutions; \cite{Alonso2019}, furthermore, a systematic review of work on LA dashboards calls for more empirical user-centered LA system development \cite{Matcha2020}. This study was guided by two questions: i) What are key learners' expectations from learning analytics? and ii) How can we implement these expectations in learning analytics-based solution with a focus on self-regulated learning? An emphasis on a user-centered approach to LA has guided this research. We introduce a LA system with the goal of offering students a checkpoint with a prediction of their final academic outcome during the semester. There are multiple recognized benefits of such an interface; firstly, students obtain feedback on their learning patterns; secondly, the results of timely prediction are explained and offer students enough time to self-regulate their performance and lastly, no additional manpower from lecturers is needed to offer personalized learning suggestions and corrective strategies to students. Self-regulated processes are “the processes whereby students activate and sustain cognitions, affects, and behaviors that are systematically oriented toward the attainment of personal goals” \cite{schunk2011handbook}. A systematic review of self-regulated learning research based in connection to learning analytics which has been conducted by \cite{Viberg2020} showed that potential of LA to improve learning outcomes had been discussed in very few studies. Most existing solutions offer no support, while some solutions offer visualizations and feedback to learners. Recommendations based on learning analytics evidence were used only in two studies. Interface application was limited to one subject, though the project is scalable. The contributions of this work are three-fold: i) The translation of the SELAQ questionnaire in Slovenian was presented along with the results gathered in Slovenian higher education institution, offering empirical insight into student expectations for learning analytics and some comparison with other countries. ii) Secondly, qualitative data was gathered to complement the understanding of learners' needs for LA services. iii) Finally, a solution proposal was presented based on the gathered requirements. 


\section{Background and related work}

\subsection{Students' expectations of Learning Analytics}
Student expectations as stakeholders have been observed in different contexts. Whitelock-Wainwright \cite{SELAQ} measured students' expectations of LA and proposed a 12-item instrument for the measurement of two core expectation categories - service and privacy expectations. The study focused on students and left out other LA stakeholders. The proposed instrument and its translation were later validated in Estonian, Dutch, and Spanish higher education settings \cite{SELAQ-multilang}. Researchers have used it as a study tool to assess stakeholders' expectations of learning analytics in European \cite{Engström2022} and Latin American \cite{HILLIGER2020100726} higher education institutions. Tsai et al. \cite{Tsai2020} used surveys and focus groups to identify students' expectations of privacy along with their behavior regarding privacy. Researchers highlighted purpose, access, and anonymity as key benchmarks of ethics and privacy. Furthermore, transparency and communication were recognized as key bases for LA adoption.

\subsection{Learning analytics and self-regulated learning}
Learning analytics dashboards are “single displays that aggregate different indicators about learner(s), learning process(es), and/or learning context(s) into one or multiple visualizations. "\cite{Schwendimann2017}. A systematic review of the LA dashboard creation \cite{Jivet2018} showed the majority of dashboards (75\%) are developed for teachers, and less focus is put on solutions targeted at learners. The review recognized the investigation of particular requirements of user groups as an open issue in the field. Additionally, only two observed propositions provided feedback or warnings to users, and only four papers used multiple data sources, indicating this as an opportunity for future research. LA for self-regulated learning consists of: i) a calculation of student behavior and ii) a recommendation for changes about how learning is carried out and how to change it \cite{winne_learning_2017}. A systematic review of LA studies for self-regulated learning \cite{Viberg2020} showed that existing research mainly focuses on calculations, which do not automatically lead to improvements in learning. In this sense, the potential of LA has not been explored to a full degree. 

Related work on predicting students’ course achievement used indicators of regular study, submission of assignments, the number of sessions, proof of reading the course \cite{YOU201623}, logs from virtual learning environments \cite{Wang2022} along with demographic data \cite{Kuzilek2015OUAA}, and grades \cite{AlAzawei2020} in their prediction models. Explanations of the model's predictions have been introduced lately, with \cite{Alonso2019} offering verbal explanations (i.e. "Evaluation is Pass because the number of assessments is high, and ...").

\section{Method}
This research followed the initial steps of the intelligent user interface (IUI) development process as defined by \cite{Gonclaves2019}, who recognized three sub-processes (i) IUI analysis and requirements, (ii) IUI design and implementation, and (iii) IUI verification and validation. In this paper, we focus on the first two steps of the development process. Users' expectations were analyzed with the Student expectations of learning analytics questionnaire (SELAQ) \cite{SELAQ} and further defined with focus groups.

\subsection{Quantitative analysis of students' expectations}
As conducted in related work\cite{SELAQ} \cite{SELAQ-multilang} , responses to the items were set on two 7‐point Likert scales (1 = Strongly Disagree; 7 = Strongly Agree), corresponding to ideal and predicted expectations, where ideal expectations are focused on what students would ideally like to happen and predicted expectations are focused on what students expect and believe would happen in reality. The original 12-item questionnaire was translated by a researcher who is a native Slovene speaker. After the initial translation, another researcher, also a native speaker, determined if the original meaning of the questionnaire items was preserved. If any disparities from the English questionnaire were discovered, the researchers made minor adjustments to the translation to better align the item wordings with the original SELAQ. The finalized translation is available in the attached in supporting materials of this paper. A short introduction about what LA is was also added to the survey. To improve the signal-to-noise ratio of questionnaire responses, 10\% of the fastest responses were excluded from the analysis as per \cite{Berger2021-gg} suggestion. As the translation of this questionnaire has not been validated before, the factor structure of the questionnaire was analyzed with exploratory factor analysis to observe if any underlying factors occurred due to the translation.

\subsection{Qualitative analysis of students' expectations}
The methodological combination of surveys, followed by focus groups, is widely used and was proposed by \cite{morgan1996focus}. In this setting, a focus group is the primary method of data collection, while surveys provide preliminary data that guides their application. Focus groups allow in-depth exploration of students' expectations without the in-advance set limitations of questionnaires. A semi-structured interview approach under the informed consent of the participants was used, with questions set around students' perceptions and expectations about LA in higher education. In the introduction part, a short definition of the LA field was presented to the participants. To promote further discussion, all focus groups were presented with the next set of five questions: i) What information would you be interested in regarding LA and your studies? ii) How often would you like to have access to the information? Would you like to receive reminders or keep on-demand access? iii) Would you like to have insight into LA yourself? Would you like teaching staff to have access as well? Would you like them to carry out interventions if your metrics reach marginal values? iv) How do you feel about data collection, processing, and publication in view of the LA solution? v) What contribution of learning analytics do you recognize for you and your studies? Probes were utilized in addition to the above questions to promote deeper explanations. The objective was to increase the number of focus groups that achieved saturation, i.e. until no new themes emerge or themes start repeating \cite{Krueger1998}. All students taking part in the SELAQ questionnaire were invited to participate in focus groups during the course of one week. No additional incentive was provided besides complimentary coffee and snacks. Four focus groups were formed with a total of 19 participants, who were divided into groups based on their availability. The group size variability was, therefore, unavoidable, varying between 3 and 7 students. Sessions were anonymous, and no information about participants was gathered apart from their year of study. Sessions were recorded (audio). Transcriptions of all focus groups were conducted after the meetings. Each audio file was translated verbatim, cleared of pauses and any noise, and prepared for content analysis in a text file. Anonymity was maintained with pseudonyms.

\section{Results}

\subsection{SELAQ results}
Groups of students from five different IT-oriented study programs at the bachelor, master, and PhD levels at \emph{affiliated faculty} were included in this study. The questionnaire was delivered in an online form during the pedagogical process, where students were physically present. Together, 306 responses were obtained, with 276 remaining after the noise reduction. Most respondents were male (n=170), 92 were female, remaining 14 refused to answer this question or responded with another. A third (n=114) of the respondents were first-year bachelor's students, 82 were second-year, and 20 were third-year bachelor's students. Further, 46 were first-year master's students, 6 were second-year master's students, and four were PhD students. Four students did not wish to answer this question. The mean respondent's age was 20.5 years (SD=1.9, Me=20), the youngest respondent was 18 years old, and the oldest was 33, 4 respondents did not wish to reveal their ages. The mean values of SELAQ questionnaire items are presented in Figure \ref{fig:SELAQ}, where it is visible that the average responses are higher on the ideal compared to the predicted expectation scale. Generally, the mean values of ethical and privacy items (ranging between 5.72 and 6.39) are higher compared to service feature items (ranging between 5.45 and 5.98). 

Exploratory factor analysis (EFA) extraction with principal axis factoring and oblimin rotation was conducted separately for the questions for ideal and predicted expectations. Observing the 12 questionnaire items for measuring ideal expectations, EFA revealed all items correlated at least .3 with at least one other item, suggesting reasonable factorability. Secondly, the Kaiser-Meyer-Olkin measure of sampling adequacy was .884 (above the recommended value of .8), and Bartlett’s test of sphericity was significant ($\chi^2$ (276) = 1267.025, p < .000). The communalities were all above .3, further confirming that each item shared some common variance with other items. Two factors were discovered, cumulatively explaining 46.57\% of the variance. Similarly, the EFA of predicted expectations revealed all items correlated at least .3 with at least one other item, the Kaiser-Meyer-Olkin measure of sampling adequacy was .916, and Bartlett’s test of sphericity was significant ($\chi^2$ (276) = 1519.18, p < .001). The communalities were all above .3, further confirming that each item shared some common variance with other items. Given the above indicators, factor analysis was deemed to be suitable for all questionnaire items. Two factors were discovered, cumulatively explaining 53.74\% of the variance. Discovered factor structure for both ideal and predicted expectations, aligned with the SELAQ's structure; Q1-Q3 and Q5-Q6 relate to ethical and privacy expectations and Q4 and Q7-Q12 to service feature expectations as shown in Table \ref{tab:SELAQ-matrix}.

  \begin{minipage}{\textwidth}
  \begin{minipage}[b]{0.47\textwidth}
      \centering
      \includegraphics[width=1\linewidth]{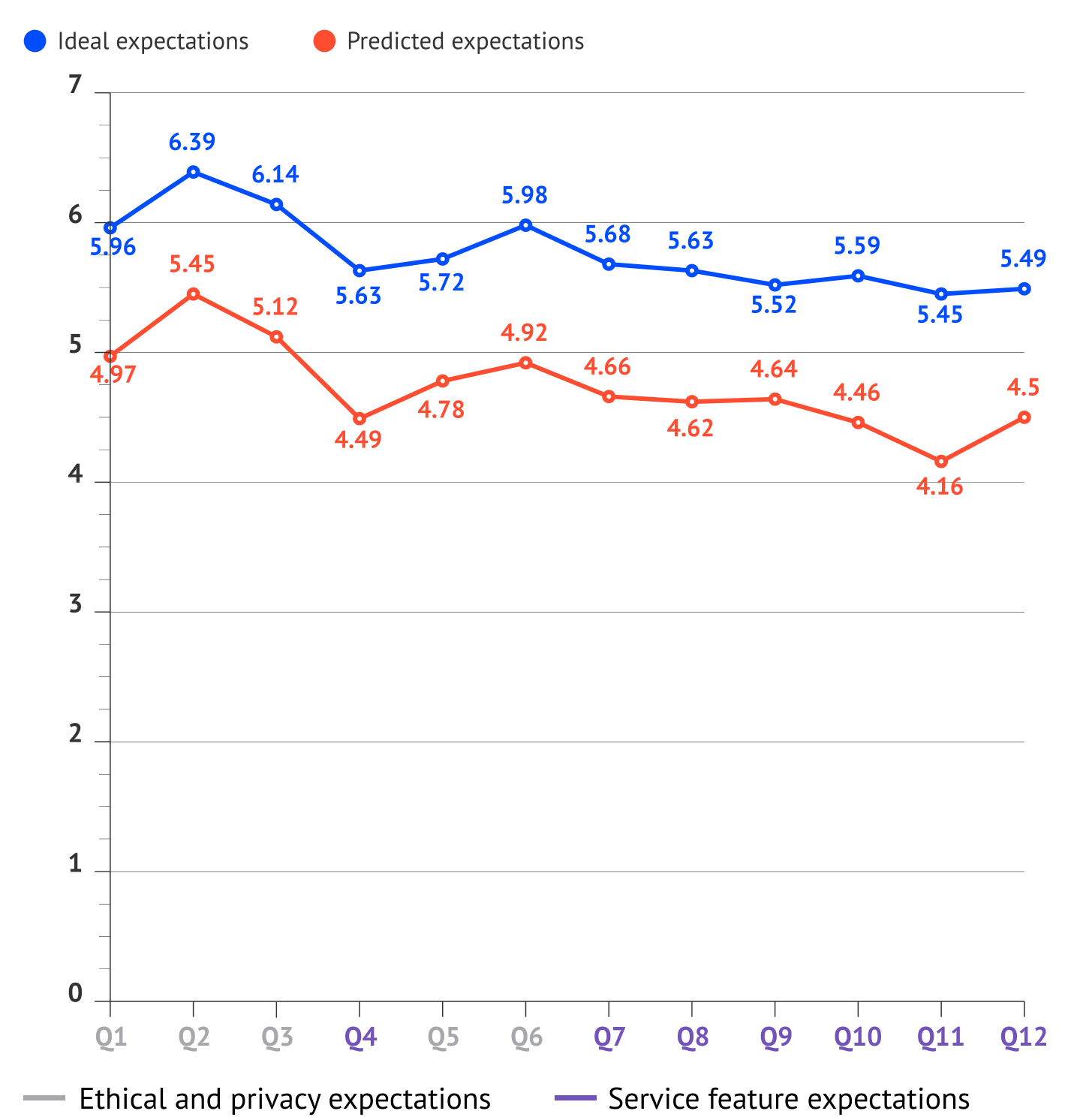}
      \captionof{figure}{Mean values for Student expectations of learning analytics questionnaire items by expectation type}
    \label{fig:SELAQ}
    \Description{Line graph of SELAQ values, displaying separate line for ideal and predicted expectations. Graph's Y axis has 7 values as per Likert scale, while the X axis hold 12 questions. Questions 1-3 and 5-6 are marked with another color as they represent items related to ethical expectations, while other questions relate to service features. Values reported for ideal expectations are as follows: 5.96, 6.39, 6.14, 5.63, 5.72, 5.98, 5.68, 5.63, 5.52, 5.59, 5.45, 5.49. Values reported for predicted expectations are 4.97, 5.45, 5.12, 4.49, 4.78, 4.92, 4.66, 4.62, 4.64, 4.46, 4.16, 4.50; and correlate with SELAQ question numbers.}

\end{minipage}
\small
  \hfill
  \begin{minipage}[b]{0.45\textwidth}
    \label{tab:SELAQ-matrix}
\begin{tabular}{lrrrr}

\toprule
 & \multicolumn{2}{c|}{Ideal} & \multicolumn{2}{c}{Predicted} \\
 \midrule
 & \multicolumn{1}{c}{Factor 1} & \multicolumn{1}{c|}{Factor 2} & \multicolumn{1}{c}{Factor 1} & \multicolumn{1}{c}{Factor 2} \\
\midrule
Q1 & -0,050 & \multicolumn{1}{r|}{-0,654} & 0,069 & 0,593 \\
Q2 & 0,011 & \multicolumn{1}{r|}{-0,730} & -0,060 & 0,781 \\
Q3 & 0,032 & \multicolumn{1}{r|}{-0,682} & -0,032 & 0,722 \\
Q4 & 0,631 & \multicolumn{1}{r|}{-0,016} & 0,414 & 0,312 \\
Q5 & 0,238 & \multicolumn{1}{r|}{-0,486} & 0,003 & 0,762 \\
Q6 & 0,006 & \multicolumn{1}{r|}{-0,773} & 0,097 & 0,712 \\
Q7 & 0,540 & \multicolumn{1}{r|}{-0,269} & 0,572 & 0,230 \\
Q8 & 0,660 & \multicolumn{1}{r|}{-0,060} & 0,676 & 0,073 \\
Q9 & 0,631 & \multicolumn{1}{r|}{-0,021} & 0,750 & -0,026 \\
Q10 & 0,627 & \multicolumn{1}{r|}{-0,090} & 0,686 & 0,013 \\
Q11 & 0,715 & \multicolumn{1}{r|}{0,166} & 0,845 & -0,121 \\
Q12 & 0,621 & \multicolumn{1}{r|}{-0,074} & 0,755 & 0,002 \\
\bottomrule
\end{tabular}
 \captionof{table}{SELAQ Pattern matrices, \\ Principal Axis Factoring}
\end{minipage}
\end{minipage}

\subsection{Focus group results}
Transcribed audio from focus groups was coded; the scheme used included five first-level codes; service features (students' expectations for features included in the virtual learning environment), feedback and notifications (students' requirements for timely and individualized feedback on their learning progress and channels for receiving this information), ethics and privacy (challenges related to data acquisition, processing, security, and anonymization), access (considering various stakeholders and their access to LA data) and perceived LA advantages (perceived usefulness of LA for students). Analysis of codes used in transcription showed service features were the most often mentioned theme with 55\% of coding references, followed by feedback and notifications (17.3\%), access (n=13.5\%), ethics and privacy (8.8\%), and perceived LA advantages (5.4\%). Each first-level code included several related second-level codes. Additionally, quotes were extracted and reviewed to complement the extracted expectations, displayed in the Table \ref{tab:focus_group_table} with information about positive or negative emotions associated with support data. 


 

\begin{table}[h]
\small
    \centering
    \begin{NiceTabular}{clll}
    \toprule
    &Extracted expectations &Descriptions &Support data \\
    \midrule
    \Block{14-1}{\rotate Service features}
    &Peer comparison &Requirements for anonymized student ranking, &25 mentions of peer comparison, \\
    & &visualizations and percentile distribution of class. &6 of percentile grade overview \\
    &Review of progress &Overview of student's progress towards their &3 mentions of comparison between  \\
    & &goals and their achievement. &current and desired grade, 7 of  \\
    & & &grade points achieved  \\
    &Historical data overview &Average grades in the last year, historical  &17 mentions of passability, 12 of \\
    & &passability percentage &historical data overview\\
    &Grade prediction &Early grade prediction and the basis for it &13 mentions of prediction \\
    &Aggregated data &Aggregated data of students' strengths &6 mentions of aggregation\\
    & &and weaknesses over multiple courses \\
    &Behaviour trends &Behaviour trends and their impact &18 mentions of behavior trends \\
    & & on academic success & \\
    &Yearly survey insight &Insight in academic survey results students take after &7 mentions of survey\\
    & & completing courses &\\
    \midrule
    \Block{13-1}{\rotate Access and permission}
    &Various access options & Options for any time and anywhere access &6 mentions of on-demand access,\\
    &and reminders &complemented with email reminders. &6 of email notifications\\
    &Notifications &Notifications at course milestones or timely intervals &12 mentions of time intervals,\\
    & &4 of course milestone notifications\\
    &Interventions &Interventions for students &4 mentions of a student group  \\
    &    & &interventions, 3 of staff interventions \\
    & Role-related permissions &Various levels of data access based on roles &12 mentions of staff access, 6  \\
   & & &mentions of peer permissions, 4\\
   & & &negative mentions of public access\\
   & & &4 mentions of data use in university \\
    &Personalization &Displaying different data based on student's &5 mentions of personalization\\
    & &psychological profile to increase their motivation &5 mentions of deception to increase,\\
    & & &motivation\\
    \midrule
    \Block{5-1}{\rotate Data, privacy}
    &Full anonymization &Full anonymization expectations &15 mentions of anonymization,\\
    & &for shared data &5 mentions of issues with student\\
    & &de-anonimization in small groups\\
    &Opt-out possibility  &Possiblity to opt-out of data collection and  &5 mentions of an opt-out \\
    & &processing as per GDPR. &option\\
    \bottomrule
    \end{NiceTabular}
    \caption{Summary of main students' expectations, extracted from focus groups }
    \label{tab:focus_group_table}
\end{table}


\subsection{User interface}
\subsubsection{Data}
Data from a mandatory introductory course in one of the bachelor's engineering study programs at \emph{affiliated faculty} was used. The course is organized in the fall semester for first-year students and is attended by around 100 students. Students can obtain between 0-100 grade points, which are later categorized in final grades between 5 and 10 (grade points 0-49 are assigned grade 5, 50-59 grade 6, 60-69 grade 7, 70-79 grade 8, etc.). The threshold for a positive grade is set at 50 grade points. The final grade is calculated from eight assignments (35 grade points), two quizzes (15 grade points), two midterms (15 grade points each), and an oral exam (20 grade points). Students must also obtain at least a passing grade (25 grade points) from assessments and quizzes combined. The average grade for the course in the academic year 2021/22 was 7.8; 18 out of 106 students failed the course. Anonymized log of interactions with the Moodle virtual learning environment, demographic data, and grades from the fall semester of the academic year 2021/22 were used as data sources for predictions. 

\subsubsection{Architecture and prediction models}
The user interface was developed using HTML and Javascript. The grade prediction model was prepared in Python version 3.8.5 in Jupyter Notebook version 6.1.4. The model was created using Numpy (version 1.19.2) and Scikit-learn (version 1.1.2). The prediction model was exposed through an API via the Flask library. Four prediction models were prepared to offer monthly updated grade predictions during the semester (October-January). 

\begin{figure}[h]
    \centering
    \includegraphics[width=0.55\textwidth]{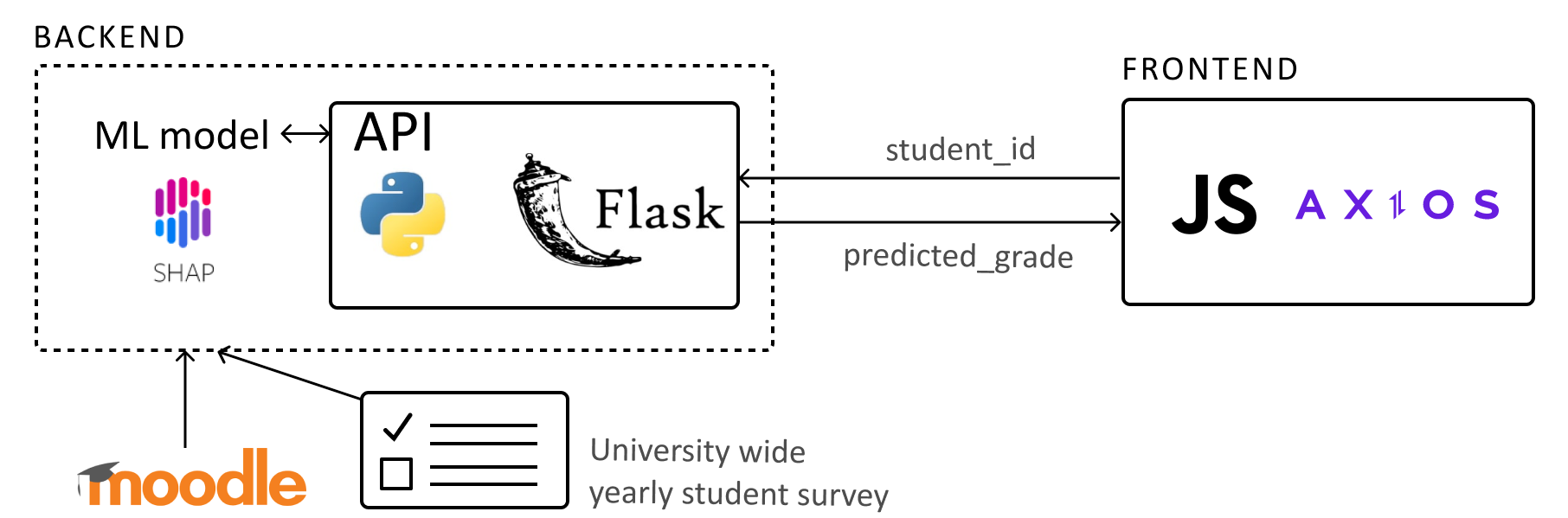}
    \caption{The architecture of the proposed solution with data sources}
    \Description[architecture sketch]{Rectangle for backend includes ML model, SHAP library logo, and an inner rectangle, displaying API includes Python logo and Flask library logo. API Rectangle is connected with the frontend rectangle, which is filled with logos of JavaScript and Axios. Two data sources are connected to backend; Moodle and University wide yearly student survey.}
    \label{fig:architecture}
\end{figure}

Final dataframe utilized demographic data (gender, disability status, schedule group), VLE data (number of clicks for each month, number of clicks for each data type, number of clicks before the start of the semester, date of the first VLE interaction, sum of all clicks) and data about available grades (tasks 1-8, midterms, quizzes 1-2). Feature suggestions were derived from related work \cite{Wang2022}\cite{Kuzilek2015OUAA}\cite{AlAzawei2020}\cite{YOU201623}. Predictions were made monthly, and each later model had more data available and therefore a higher number of features (i.e. each grade was used as a new feature). As the task grades (the sum of which presents the final grade) were used as features in the model, some level of multicollinearity could not be avoided. Grade prediction was conducted with two models i) a classification model was used to determine if a student will achieve a passing grade, and ii) for students with predicted passing grade, a regression model was used to predict the exact final grade in grade points. In both cases, multiple algorithms were compared, and the best performing one was selected for the model; Random Forest Classifier performed best for the classification model (compared with Logistic Regression, K-Neighbors Classifier, Gaussian Naive Bayes, Decision Tree Classifier and Linear Discriminant Analysis) and Decision Tree Regressor performed best for regression model (compared to Linear Regression, and Random Forest Regressor). To validate the performance of our models, 5-split K-Fold cross-validation from Scikit-learn was used.

\subsubsection{Interface}
\begin{figure}
    \centering
    \begin{subfigure}[b]{0.47\textwidth}
        \centering
        \includegraphics[width=1\textwidth]{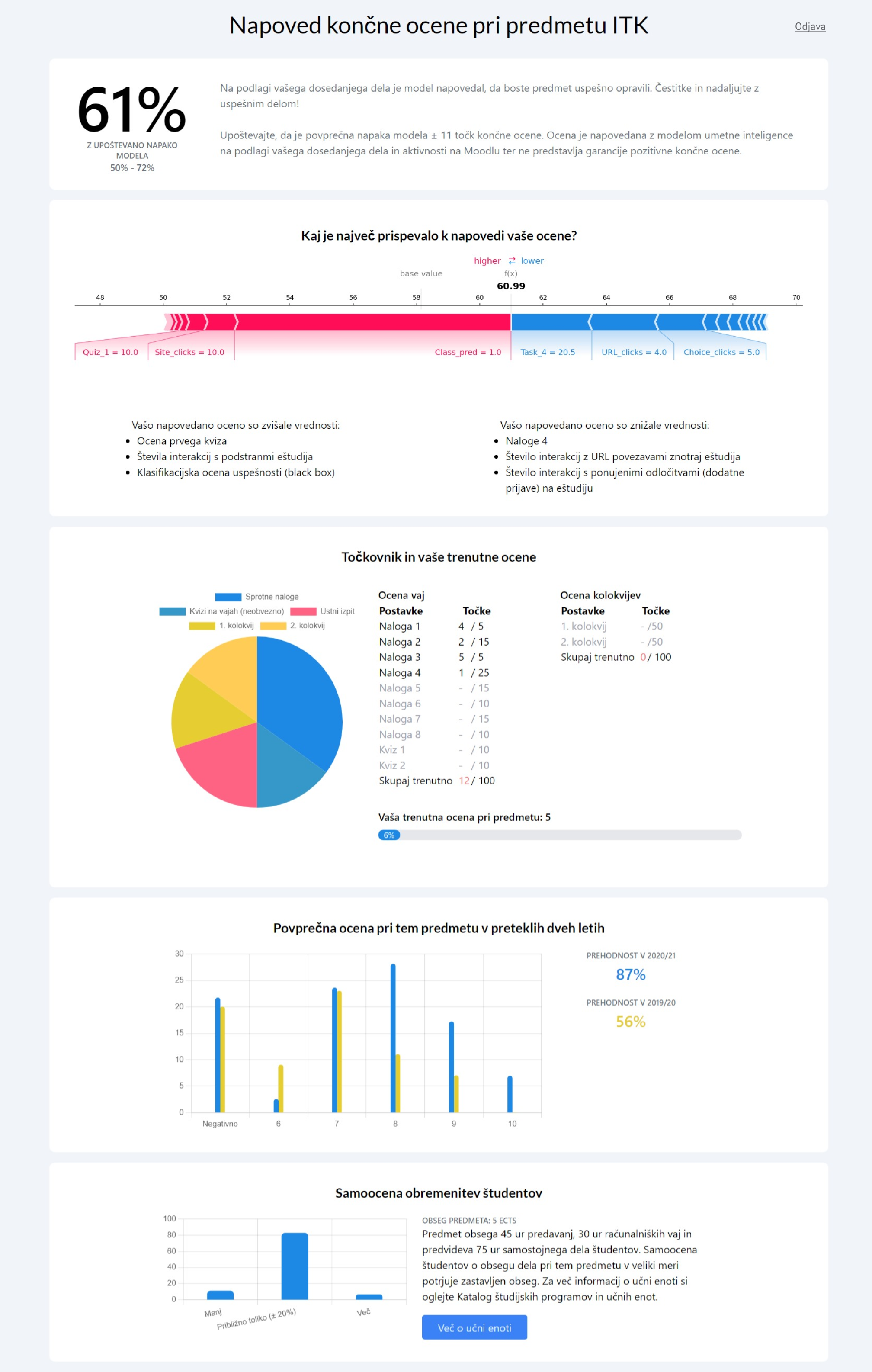} 
        \caption{Example of dashboard view for students with predicted passing grade}
        \Description[Dashboard view for students with predicted passing grade]{Consists of predicted grade, SHAP prediction explanation, textual prediction explanation, pie chart with grade elements, table of grade elements with their achieved grade, pill-chart of their grade progress,  bar chart with overview of historical grades from this course and information about passability. At the bottom, survey results about students' reported effort in class are displayed in bar chart.}
        \label{fig:success}
    \end{subfigure}\hfill
    \begin{subfigure}[b]{0.475\textwidth}
        \centering
        \includegraphics[width=1\textwidth]{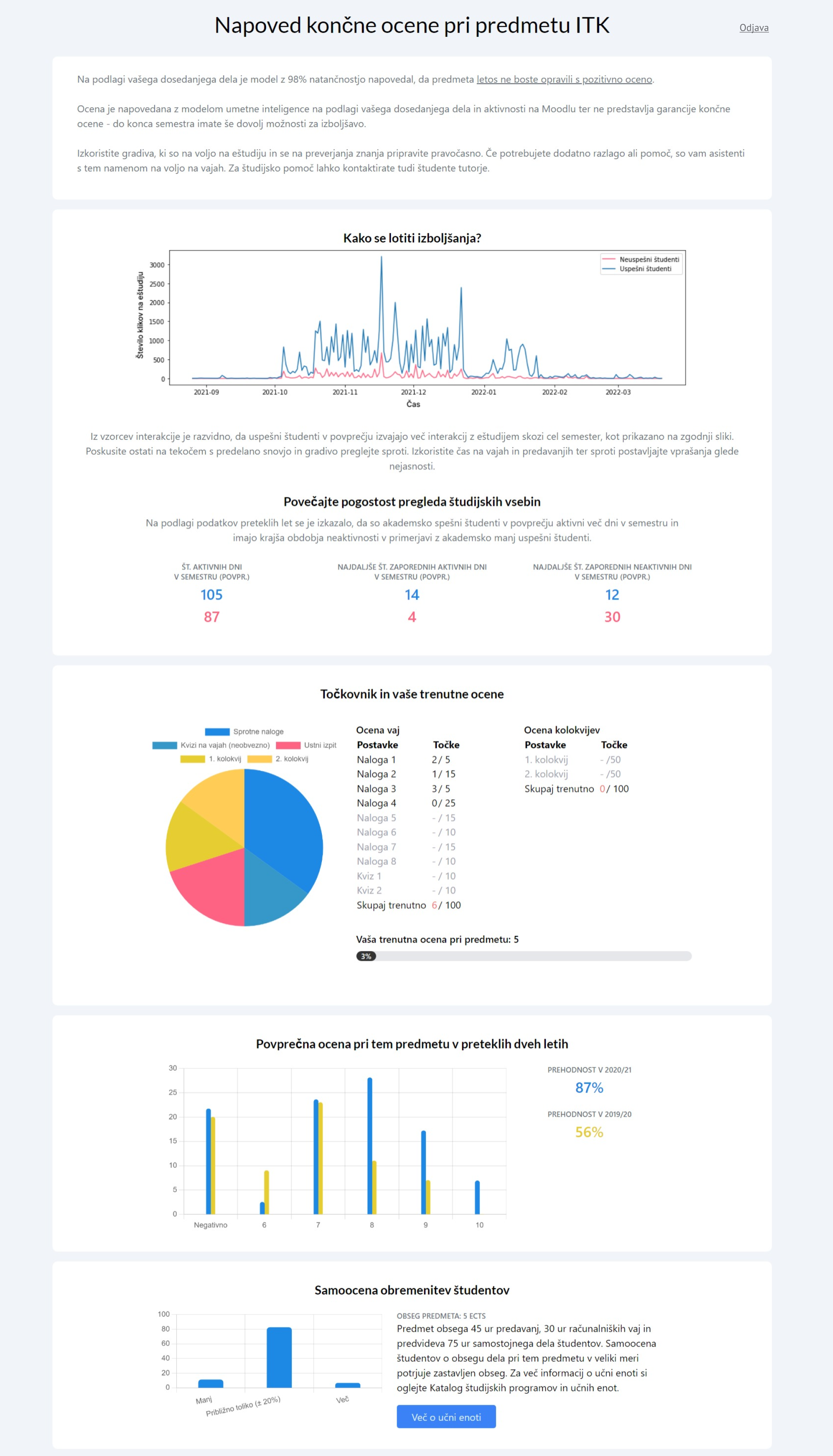} 
        \caption{Example of dashboard view for students at-risk of failing the course}
        \Description[Dashboard view for students with predicted failing grade]{Consists of description of predicted failing grade, prediction explanation, and information about available resources. This is followed by section about improving the behaviour, including line chart of average interactions in VLE, behavior described in numbers (nr. of active, inactive day, etc.), followd by pie chart with elements of the grade for this course, table of grade elements with their achieved grade, pill-chart of their grade progress, bar chart with overview of historical grades from this course and information about passability. At the bottom, survey results about students' reported effort in class are displayed in bar chart.}
        \label{fig:fail}
    \end{subfigure}
     \caption{Preview of the proposed LA interface}
\end{figure}

Following the grade prediction model, two user flows were considered in user interface development. The first flow, displayed in Figure \ref{fig:success} covered positive grade predictions, presenting students with the predicted grade, along with elements of explainable artificial intelligence in the form of visual and text explanations of the prediction. SHAP library \cite{SHAP} is utilized for feature explanation in predictions. The second user flow, presented in Figure \ref{fig:fail} shows display for students with negative grade predictions, who would be presented with a textual explanation of the prediction model and comparison of the behavior trends of academically successful and unsuccessful students, prompting learners to change their behavior. Behavior pattern display included an overview of the number of clicks during the semester, the number of active and inactive days in the virtual learning environment, and a number of consecutive active and inactive days. Some general guidelines of channels available for academic help were also provided. An overview of learner's currently obtained grades, courses grading system, and historical data on grades and course passability are presented in both cases. At last, self-reported effort in hours spent working for this course are presented; this data was taken from publicly available university-wide student surveys. 

A preliminary interface evaluation was conducted with ten students. Learners received access to the dashboard with anonymized data of two representative users; one academically successful and one at risk of failing the course. After interaction with the LA dashboard, their feedback was gathered with a questionnaire consisting of System usability scale (SUS) questions and questions regarding privacy and feature satisfaction, where students reported their answers on a 5-point Likert scale. A SUS score of 76.5 was achieved, which is interpreted as good by the SUS evaluation key. Students have furthermore expressed their support for the use of their data in the implementation of LA (n=4.6 points on the Likert scale, SD=0.5), and reported feeling enough informed about the use and processing of their data (n=4.1, SD=0.9). Students believed that such display would, at least to some degree, motivate them in their studies (n=3.8, SD=1) and offer some help in planning their study activities (n=3.7, SD=0.8).

\section{Discussion}
This study explored the student's expectations of LA in higher education institution in Slovenia and offered a possible solution based on the gathered expectations. Observing learners' expectations, we see some similarities with findings from related work \cite{SELAQ-multilang}\cite{Engström2022} \cite{Garcia2021}, Slovenian students' ideal expectations of LA are lower compared to their predicted realistic expectations. The largest difference (1.29 points on the Likert scale) is observed in item 11, which is focusing on the obligation of teaching staff to act if students are at risk of failing and underperforming. Similar low  results for item 11 were reported in related work for Dutch, Estonian, Spanish, \cite{SELAQ-multilang} and Swedish \cite{Engström2022} samples. This item was further discussed in focus groups. Observing ethical and privacy expectation items (Items 1-5 and 6), the highest mean value was observed for Item 2, which focused on the secure keeping of the student's data. This item also reached the highest overall mean value for both ideal (M=6.39, SD=1.51) and predicted (M=5.45, SD=1.51) expectations, indicating that students highly value the security of their educational data. A low difference between ideal and predicted expectations is observed for this item (0.94 points on the Likert scale), showing the students still have some doubts about the storage and security of their data.  Observing service feature expectation items (Items 4 and 6-12), the highest mean score can be observed for item 6 (the university will require further consent if my data will be used for a purpose differing from what was originally stated), indicating the importance of control students wish to have over the processing of their data. The mean value of this item is high for ideal (M=5.98, SD=1.54) and predicted (M=4.92, SD=1.54) expectations. A low mean score can be observed in Item 4 (the university will keep me informed about my learning progress), possibly indicating a low perceived need to receive such updates or low expectation for their delivery. The lowest standard deviation between of 1.33 points was observed in Item 2 (ideal expectations for strong data security), while the highest standard deviation of 1.67 points on the Likert scale was observed for Item 11 (Ideal expectation for teaching staff obligation to take action), with later indicating different opinions of students about the staff action based on learning analytics. The EFA conducted on SELAQ data confirmed the questionnaire's two-factor structure, which has been preserved through translation as already suggested by \cite{SELAQ-multilang} with Dutch and Estonian translations.

Analyzing data from focus groups, we observed learners point out peer comparison, historical data (their own history and statics of previous students), grade prediction (early in the semester), and behavior trends overview (specifically, the behavior of academically successful students) as some of the most important expectations of the learning analytics systems. Aggregated overview of multiple courses and personalization based on learners' psychological profiles were also often mentioned. With regard to privacy and data processing, full anonymization was pointed out in all groups as an important system characteristic. Some doubts about the extent of anonymization were voiced in cases of small groups of students, where de-anonymization of individuals from statistics is possible even with limited known data points. Three groups recognized the possibility to opt-out of the data use for statistics as an important feature. In terms of access and intervention based on learning analytics data, students felt comfortable sharing generalized analytical data with educators directly involved in each course but were less inclined to share data with other teaching staff not directly involved with each course. Public access to a statistical overview of anonymized data was negatively viewed by students from three focus groups. Students viewed teaching staff interventions as less desirable but were open to the idea of interventions focused on entire groups of students. Surprisingly, students recognized increased motivation for higher and average-performing students as the result of access to LA, while decreased motivation was predicted for underperforming students. Learners recognized elements of self-regulated learning as one of the primary advantages of LA. On-demand access combined with notifications in time intervals (either monthly or at course milestones) was pointed out as desirable. Personalization in terms of content adaptation was desirable; interestingly multiple students suggested deception to increase the motivation of students at-risk of underperforming. 

Gathered student requirements complement the existing body of research about learners' expectations of LA \cite{HILLIGER2020100726} \cite{SELAQ-multilang} \cite{Engström2022}. At this point, only a subset of gathered requirements was actually implemented in the proposed solution, as some features were related to large-scale implementation (e.g. data aggregation) or were not directly related to the LA solution (e.g. teacher's interventions). 



\section{Limitations and future work}
\subsection{Limitations}
This study focused solely on the expectations of higher education students in selected Slovenian higher education institution. Generalizations of this findings are limited. This study did not consider other stakeholders, such as teachers and faculty staff. The selected SELAQ questionnaire has been used in multiple countries. Nevertheless, the proposed Slovenian translation has not been used before and has not been statistically validated in this study using the approach previously proposed by \cite{SELAQ-multilang}. Furthermore, the SELAQ questionnaire offers only a general understanding of what students expect of LA services. The questionnaire is focused on student expectations; the view of the faculty and teaching staff is not considered in this approach, though they play an important role in the adaptation of LA. 

We acknowledge that the use of focus groups is a qualitative approach and as such has its limitations. The selected sample of participants might not be representative of the whole target group of students. Due to sampling, the applicability of this study is limited to selected Slovenian higher education setting, where the research was conducted in a narrower view. The moderators aimed to remain objective and reduce their effect on participants; however, it is impossible to reduce their impact completely. Moderator's skills can also affect responses and skew the results, which poses a concern in this study as only one researcher acted as the moderator. The effect of attitude polarization on participants is also a known issue, with attitudes becoming more extreme after the discussion in a group \cite{morgan1996focus}. Some bias can also be expected as respondents might feel peer pressure to give similar answers as other participants.

The proposed interface was currently developed only for one course, therefore not yet fully meeting the students' recognized requirements. Further models are needed in order to expand the interface to other courses and account for year-to-year structural course changes. The prediction model was fitted with historical data from one academic year. Any unexpected  deviations in students' behavior might negatively affect the accuracy of grade predictions. Cross-validation was used to avoid overfitting. Furthermore, the largest issue related to the model is false positive predictions, i.e. cases where students are falsely presented with passing grade predictions. Currently, only the preliminary evaluation was conducted, which offered limited insight into system usability and no insight into student impact in practice. 

\subsection{Future work}
An ethical board review of the proposed interface is planned, as presenting students with predicted grades has been reported \cite{Lim2021} to potentially trigger feelings of anxiety, stress, guilt, and frustration in some cases of students; however, others experienced relief about doubting their abilities and relief of anxiety. Experiment for evaluation of the proposed user interface is planned, following the ethical approval and guidelines. Further on, additional grade prediction models tailored for other courses should be added to reach the expectations of subject coverage. Furthermore, before the LA application on a wider scale, the expectations of teachers and faculty staff should be analyzed and considered.


\bibliographystyle{ACM-Reference-Format}
\bibliography{sample-base}


\begin{thebibliography}{25}


\ifx \showCODEN    \undefined \def \showCODEN     #1{\unskip}     \fi
\ifx \showDOI      \undefined \def \showDOI       #1{#1}\fi
\ifx \showISBNx    \undefined \def \showISBNx     #1{\unskip}     \fi
\ifx \showISBNxiii \undefined \def \showISBNxiii  #1{\unskip}     \fi
\ifx \showISSN     \undefined \def \showISSN      #1{\unskip}     \fi
\ifx \showLCCN     \undefined \def \showLCCN      #1{\unskip}     \fi
\ifx \shownote     \undefined \def \shownote      #1{#1}          \fi
\ifx \showarticletitle \undefined \def \showarticletitle #1{#1}   \fi
\ifx \showURL      \undefined \def \showURL       {\relax}        \fi
\providecommand\bibfield[2]{#2}
\providecommand\bibinfo[2]{#2}
\providecommand\natexlab[1]{#1}
\providecommand\showeprint[2][]{arXiv:#2}

\bibitem[Al-Azawei and Al-Masoudy(2020)]%
        {AlAzawei2020}
\bibfield{author}{\bibinfo{person}{Ahmed Al-Azawei} {and}
  \bibinfo{person}{Miami Al-Masoudy}.} \bibinfo{year}{2020}\natexlab{}.
\newblock \showarticletitle{Predicting Learners' Performance in Virtual
  Learning Environment (VLE) based on Demographic, Behavioral and Engagement
  Antecedents}.
\newblock \bibinfo{journal}{\emph{International Journal of Emerging
  Technologies in Learning (IJET)}} \bibinfo{volume}{15}, \bibinfo{number}{9}
  (\bibinfo{year}{2020}), \bibinfo{pages}{60--75}.
\newblock


\bibitem[Alonso and Casalino(2019)]%
        {Alonso2019}
\bibfield{author}{\bibinfo{person}{Jos{\'e}~M. Alonso} {and}
  \bibinfo{person}{Gabriella Casalino}.} \bibinfo{year}{2019}\natexlab{}.
\newblock \showarticletitle{Explainable Artificial Intelligence for
  Human-Centric Data Analysis in Virtual Learning Environments}. In
  \bibinfo{booktitle}{\emph{Higher Education Learning Methodologies and
  Technologies Online}}, \bibfield{editor}{\bibinfo{person}{Daniel Burgos},
  \bibinfo{person}{Marta Cimitile}, \bibinfo{person}{Pietro Ducange},
  \bibinfo{person}{Riccardo Pecori}, \bibinfo{person}{Pietro Picerno},
  \bibinfo{person}{Paolo Raviolo}, {and} \bibinfo{person}{Christian~M.
  Stracke}} (Eds.). \bibinfo{publisher}{Springer International Publishing},
  \bibinfo{address}{Cham}, \bibinfo{pages}{125--138}.
\newblock
\showISBNx{978-3-030-31284-8}


\bibitem[Berger and Kiefer(2021)]%
        {Berger2021-gg}
\bibfield{author}{\bibinfo{person}{Alexander Berger} {and}
  \bibinfo{person}{Markus Kiefer}.} \bibinfo{year}{2021}\natexlab{}.
\newblock \showarticletitle{Comparison of Different Response Time Outlier
  Exclusion Methods: A Simulation Study}.
\newblock \bibinfo{journal}{\emph{Front Psychol}}  \bibinfo{volume}{12}
  (\bibinfo{date}{June} \bibinfo{year}{2021}), \bibinfo{pages}{675558}.
\newblock


\bibitem[Engström et~al\mbox{.}(2022)]%
        {Engström2022}
\bibfield{author}{\bibinfo{person}{Linda Engström}, \bibinfo{person}{Olga
  Viberg}, \bibinfo{person}{Olle Bälter}, {and} \bibinfo{person}{Stefan
  Hrastinski}.} \bibinfo{year}{2022}\natexlab{}.
\newblock \showarticletitle{Students’ Expectations of Learning Analytics in a
  Swedish Higher Education Institution}. In \bibinfo{booktitle}{\emph{2022 IEEE
  Global Engineering Education Conference (EDUCON)}}.
  \bibinfo{publisher}{IEEE}, \bibinfo{address}{Tunis, Tunisia},
  \bibinfo{pages}{1975--1980}.
\newblock
\urldef\tempurl%
\url{https://doi.org/10.1109/EDUCON52537.2022.9766482}
\showDOI{\tempurl}


\bibitem[Garcia et~al\mbox{.}(2021)]%
        {Garcia2021}
\bibfield{author}{\bibinfo{person}{Samantha Garcia}, \bibinfo{person}{Elaine
  Cristina~Moreira Marques}, \bibinfo{person}{Rafael~Ferreira Mello},
  \bibinfo{person}{Dragan Ga{\v{s}}evi{\'{c}}}, {and}
  \bibinfo{person}{Taciana~Pontual Falc{\~a}o}.}
  \bibinfo{year}{2021}\natexlab{}.
\newblock \showarticletitle{Aligning Expectations About the Adoption of
  Learning Analytics in a Brazilian Higher Education Institution}. In
  \bibinfo{booktitle}{\emph{Artificial Intelligence in Education}},
  \bibfield{editor}{\bibinfo{person}{Ido Roll}, \bibinfo{person}{Danielle
  McNamara}, \bibinfo{person}{Sergey Sosnovsky}, \bibinfo{person}{Rose Luckin},
  {and} \bibinfo{person}{Vania Dimitrova}} (Eds.). \bibinfo{publisher}{Springer
  International Publishing}, \bibinfo{address}{Cham},
  \bibinfo{pages}{173--177}.
\newblock
\showISBNx{978-3-030-78270-2}


\bibitem[Gon\c{c}alves and da~Rocha(2019)]%
        {Gonclaves2019}
\bibfield{author}{\bibinfo{person}{Taisa~Guidini Gon\c{c}alves} {and}
  \bibinfo{person}{Ana Regina~Cavalcanti da Rocha}.}
  \bibinfo{year}{2019}\natexlab{}.
\newblock \showarticletitle{Development Process for Intelligent User
  Interfaces: An Initial Approach}. In \bibinfo{booktitle}{\emph{Proceedings of
  the XVIII Brazilian Symposium on Software Quality}} (Fortaleza, Brazil)
  \emph{(\bibinfo{series}{SBQS'19})}. \bibinfo{publisher}{Association for
  Computing Machinery}, \bibinfo{address}{New York, NY, USA},
  \bibinfo{pages}{210–215}.
\newblock
\showISBNx{9781450372824}
\urldef\tempurl%
\url{https://doi.org/10.1145/3364641.3364665}
\showDOI{\tempurl}


\bibitem[Hilliger et~al\mbox{.}(2020)]%
        {HILLIGER2020100726}
\bibfield{author}{\bibinfo{person}{Isabel Hilliger}, \bibinfo{person}{Margarita
  Ortiz-Rojas}, \bibinfo{person}{Paola Pesántez-Cabrera},
  \bibinfo{person}{Eliana Scheihing}, \bibinfo{person}{Yi-Shan Tsai},
  \bibinfo{person}{Pedro~J. Muñoz-Merino}, \bibinfo{person}{Tom Broos},
  \bibinfo{person}{Alexander Whitelock-Wainwright}, {and} \bibinfo{person}{Mar
  Pérez-Sanagustín}.} \bibinfo{year}{2020}\natexlab{}.
\newblock \showarticletitle{Identifying needs for learning analytics adoption
  in Latin American universities: A mixed-methods approach}.
\newblock \bibinfo{journal}{\emph{The Internet and Higher Education}}
  \bibinfo{volume}{45} (\bibinfo{year}{2020}), \bibinfo{pages}{100726}.
\newblock
\showISSN{1096-7516}
\urldef\tempurl%
\url{https://doi.org/10.1016/j.iheduc.2020.100726}
\showDOI{\tempurl}


\bibitem[Jivet et~al\mbox{.}(2018)]%
        {Jivet2018}
\bibfield{author}{\bibinfo{person}{Ioana Jivet}, \bibinfo{person}{Maren
  Scheffel}, \bibinfo{person}{Marcus Specht}, {and} \bibinfo{person}{Hendrik
  Drachsler}.} \bibinfo{year}{2018}\natexlab{}.
\newblock \showarticletitle{License to Evaluate: Preparing Learning Analytics
  Dashboards for Educational Practice}. In
  \bibinfo{booktitle}{\emph{Proceedings of the 8th International Conference on
  Learning Analytics and Knowledge}} (Sydney, New South Wales, Australia)
  \emph{(\bibinfo{series}{LAK '18})}. \bibinfo{publisher}{Association for
  Computing Machinery}, \bibinfo{address}{New York, NY, USA},
  \bibinfo{pages}{31–40}.
\newblock
\showISBNx{9781450364003}
\urldef\tempurl%
\url{https://doi.org/10.1145/3170358.3170421}
\showDOI{\tempurl}


\bibitem[Krueger(1998)]%
        {Krueger1998}
\bibfield{author}{\bibinfo{person}{Richard Krueger}.}
  \bibinfo{year}{1998}\natexlab{}.
\newblock \bibinfo{title}{Moderating Focus Groups}.
\newblock
\newblock
\urldef\tempurl%
\url{https://doi.org/10.4135/9781483328133}
\showDOI{\tempurl}


\bibitem[Kuzilek et~al\mbox{.}(2015)]%
        {Kuzilek2015OUAA}
\bibfield{author}{\bibinfo{person}{Jakub Kuzilek}, \bibinfo{person}{Martin
  Hlosta}, \bibinfo{person}{Drahomira Herrmannova}, \bibinfo{person}{Zdenek
  Zdr{\'a}hal}, {and} \bibinfo{person}{Annika Wolff}.}
  \bibinfo{year}{2015}\natexlab{}.
\newblock \showarticletitle{OU Analyse: analysing at-risk students at The Open
  University}.
\newblock \bibinfo{journal}{\emph{Learning Analytics Review}}
  (\bibinfo{year}{2015}), \bibinfo{pages}{1--16}.
\newblock


\bibitem[Lim et~al\mbox{.}(2021)]%
        {Lim2021}
\bibfield{author}{\bibinfo{person}{Lisa-Angelique Lim}, \bibinfo{person}{Shane
  Dawson}, \bibinfo{person}{Dragan Gašević}, \bibinfo{person}{Srecko
  Joksimović}, \bibinfo{person}{Abelardo Pardo}, \bibinfo{person}{Anthea
  Fudge}, {and} \bibinfo{person}{Sheridan Gentili}.}
  \bibinfo{year}{2021}\natexlab{}.
\newblock \showarticletitle{Students’ perceptions of, and emotional responses
  to, personalised learning analytics-based feedback: an exploratory study of
  four courses}.
\newblock \bibinfo{journal}{\emph{Assessment \& Evaluation in Higher
  Education}} \bibinfo{volume}{46}, \bibinfo{number}{3} (\bibinfo{year}{2021}),
  \bibinfo{pages}{339--359}.
\newblock
\urldef\tempurl%
\url{https://doi.org/10.1080/02602938.2020.1782831}
\showDOI{\tempurl}
\showeprint{https://doi.org/10.1080/02602938.2020.1782831}


\bibitem[Lundberg and Lee(2017)]%
        {SHAP}
\bibfield{author}{\bibinfo{person}{Scott~M Lundberg} {and}
  \bibinfo{person}{Su-In Lee}.} \bibinfo{year}{2017}\natexlab{}.
\newblock \showarticletitle{A Unified Approach to Interpreting Model
  Predictions}.
\newblock In \bibinfo{booktitle}{\emph{Advances in Neural Information
  Processing Systems 30}}, \bibfield{editor}{\bibinfo{person}{I.~Guyon},
  \bibinfo{person}{U.~V. Luxburg}, \bibinfo{person}{S.~Bengio},
  \bibinfo{person}{H.~Wallach}, \bibinfo{person}{R.~Fergus},
  \bibinfo{person}{S.~Vishwanathan}, {and} \bibinfo{person}{R.~Garnett}}
  (Eds.). \bibinfo{publisher}{Curran Associates, Inc.},
  \bibinfo{pages}{4765--4774}.
\newblock
\urldef\tempurl%
\url{http://papers.nips.cc/paper/7062-a-unified-approach-to-interpreting-model-predictions.pdf}
\showURL{%
\tempurl}


\bibitem[Matcha et~al\mbox{.}(2020)]%
        {Matcha2020}
\bibfield{author}{\bibinfo{person}{Wannisa Matcha},
  \bibinfo{person}{Nora'ayu~Ahmad Uzir}, \bibinfo{person}{Dragan Gašević},
  {and} \bibinfo{person}{Abelardo Pardo}.} \bibinfo{year}{2020}\natexlab{}.
\newblock \showarticletitle{A Systematic Review of Empirical Studies on
  Learning Analytics Dashboards: A Self-Regulated Learning Perspective}.
\newblock \bibinfo{journal}{\emph{IEEE Transactions on Learning Technologies}}
  \bibinfo{volume}{13}, \bibinfo{number}{2} (\bibinfo{year}{2020}),
  \bibinfo{pages}{226--245}.
\newblock
\urldef\tempurl%
\url{https://doi.org/10.1109/TLT.2019.2916802}
\showDOI{\tempurl}


\bibitem[Morgan(1996)]%
        {morgan1996focus}
\bibfield{author}{\bibinfo{person}{D.L. Morgan}.}
  \bibinfo{year}{1996}\natexlab{}.
\newblock \bibinfo{booktitle}{\emph{Focus Groups as Qualitative Research}}.
\newblock \bibinfo{publisher}{SAGE Publications}, \bibinfo{address}{Thousand
  Oaks, California}.
\newblock
\showISBNx{9781506318820}
\urldef\tempurl%
\url{https://books.google.si/books?id=LxF5CgAAQBAJ}
\showURL{%
\tempurl}


\bibitem[Nouri et~al\mbox{.}(2019)]%
        {Nouri2019}
\bibfield{author}{\bibinfo{person}{Jalal Nouri}, \bibinfo{person}{Martin
  Ebner}, \bibinfo{person}{Dirk Ifenthaler}, \bibinfo{person}{Mohammed Saqr},
  \bibinfo{person}{Jonna Malmberg}, \bibinfo{person}{Mohammad Khalil},
  \bibinfo{person}{Jesper Bruun}, \bibinfo{person}{Olga Viberg},
  \bibinfo{person}{Miguel~Ángel Conde~González}, \bibinfo{person}{Zacharoula
  Papamitsiou}, {and} \bibinfo{person}{Ulf~Dalvad Berthelsen}.}
  \bibinfo{year}{2019}\natexlab{}.
\newblock \showarticletitle{Efforts in Europe for Data-Driven Improvement of
  Education – A Review of Learning Analytics Research in Seven Countries}.
\newblock \bibinfo{journal}{\emph{International Journal of Learning Analytics
  and Artificial Intelligence for Education (iJAI)}} \bibinfo{volume}{1},
  \bibinfo{number}{1} (\bibinfo{date}{Jul.} \bibinfo{year}{2019}),
  \bibinfo{pages}{pp. 8–27}.
\newblock
\urldef\tempurl%
\url{https://doi.org/10.3991/ijai.v1i1.11053}
\showDOI{\tempurl}


\bibitem[Schunk and Zimmerman(2011)]%
        {schunk2011handbook}
\bibfield{author}{\bibinfo{person}{D.H. Schunk} {and} \bibinfo{person}{B.
  Zimmerman}.} \bibinfo{year}{2011}\natexlab{}.
\newblock \bibinfo{booktitle}{\emph{Handbook of Self-Regulation of Learning and
  Performance}}.
\newblock \bibinfo{publisher}{Taylor \& Francis}, \bibinfo{address}{New York,
  NY, USA}.
\newblock
\showISBNx{9781136881664}
\urldef\tempurl%
\url{https://books.google.si/books?id=XfOYV0lwzGgC}
\showURL{%
\tempurl}


\bibitem[Schwendimann et~al\mbox{.}(2017)]%
        {Schwendimann2017}
\bibfield{author}{\bibinfo{person}{Beat~A. Schwendimann},
  \bibinfo{person}{María~Jesús Rodríguez-Triana}, \bibinfo{person}{Andrii
  Vozniuk}, \bibinfo{person}{Luis~P. Prieto}, \bibinfo{person}{Mina~Shirvani
  Boroujeni}, \bibinfo{person}{Adrian Holzer}, \bibinfo{person}{Denis Gillet},
  {and} \bibinfo{person}{Pierre Dillenbourg}.} \bibinfo{year}{2017}\natexlab{}.
\newblock \showarticletitle{Perceiving Learning at a Glance: A Systematic
  Literature Review of Learning Dashboard Research}.
\newblock \bibinfo{journal}{\emph{IEEE Transactions on Learning Technologies}}
  \bibinfo{volume}{10}, \bibinfo{number}{1} (\bibinfo{year}{2017}),
  \bibinfo{pages}{30--41}.
\newblock
\urldef\tempurl%
\url{https://doi.org/10.1109/TLT.2016.2599522}
\showDOI{\tempurl}


\bibitem[Tsai et~al\mbox{.}(2017)]%
        {SHEILA2017}
\bibfield{author}{\bibinfo{person}{Yi-Shan Tsai}, \bibinfo{person}{Maren
  Scheffel}, {and} \bibinfo{person}{Dragan Gasevic}.}
  \bibinfo{year}{2017}\natexlab{}.
\newblock \showarticletitle{SHEILA policy framework – supporting higher
  education to integrate learning analytics}. In \bibinfo{booktitle}{\emph{The
  8th International Learning Analytics and Knowledge (LAK) Conference}}.
  \bibinfo{publisher}{ACM, New York, NY, United States},
  \bibinfo{address}{Sydney}, \bibinfo{numpages}{3}~pages.
\newblock
\urldef\tempurl%
\url{https://latte-analytics.sydney.edu.au}
\showURL{%
\tempurl}
\newblock
\shownote{The 8th International Learning Analytics and Knowledge (LAK)
  Conference ; Conference date: 05-03-2018 Through 09-03-2018}.


\bibitem[Tsai et~al\mbox{.}(2020)]%
        {Tsai2020}
\bibfield{author}{\bibinfo{person}{Yi-Shan Tsai}, \bibinfo{person}{Alexander
  Whitelock-Wainwright}, {and} \bibinfo{person}{Dragan Ga\v{s}evi\'{c}}.}
  \bibinfo{year}{2020}\natexlab{}.
\newblock \showarticletitle{The Privacy Paradox and Its Implications for
  Learning Analytics}. In \bibinfo{booktitle}{\emph{Proceedings of the Tenth
  International Conference on Learning Analytics \& Knowledge}} (Frankfurt,
  Germany) \emph{(\bibinfo{series}{LAK '20})}. \bibinfo{publisher}{Association
  for Computing Machinery}, \bibinfo{address}{New York, NY, USA},
  \bibinfo{pages}{230–239}.
\newblock
\showISBNx{9781450377126}
\urldef\tempurl%
\url{https://doi.org/10.1145/3375462.3375536}
\showDOI{\tempurl}


\bibitem[Viberg et~al\mbox{.}(2020)]%
        {Viberg2020}
\bibfield{author}{\bibinfo{person}{Olga Viberg}, \bibinfo{person}{Mohammad
  Khalil}, {and} \bibinfo{person}{Martine Baars}.}
  \bibinfo{year}{2020}\natexlab{}.
\newblock \showarticletitle{Self-Regulated Learning and Learning Analytics in
  Online Learning Environments: A Review of Empirical Research}. In
  \bibinfo{booktitle}{\emph{Proceedings of the Tenth International Conference
  on Learning Analytics \& Knowledge}} (Frankfurt, Germany)
  \emph{(\bibinfo{series}{LAK '20})}. \bibinfo{publisher}{Association for
  Computing Machinery}, \bibinfo{address}{New York, NY, USA},
  \bibinfo{pages}{524–533}.
\newblock
\showISBNx{9781450377126}
\urldef\tempurl%
\url{https://doi.org/10.1145/3375462.3375483}
\showDOI{\tempurl}


\bibitem[Wang et~al\mbox{.}(2022)]%
        {Wang2022}
\bibfield{author}{\bibinfo{person}{Xinzheng Wang}, \bibinfo{person}{Bing Guo},
  {and} \bibinfo{person}{Yan Shen}.} \bibinfo{year}{2022}\natexlab{}.
\newblock \showarticletitle{Predicting the At-Risk Online Students Based on the
  Click Data Distribution Characteristics}.
\newblock \bibinfo{journal}{\emph{Scientific Programming}}
  \bibinfo{volume}{2022} (\bibinfo{year}{2022}), \bibinfo{numpages}{12}~pages.
\newblock


\bibitem[Whitelock-Wainwright et~al\mbox{.}(2019)]%
        {SELAQ}
\bibfield{author}{\bibinfo{person}{Alexander Whitelock-Wainwright},
  \bibinfo{person}{Dragan Gašević}, \bibinfo{person}{Ricardo Tejeiro},
  \bibinfo{person}{Yi-Shan Tsai}, {and} \bibinfo{person}{Kate Bennett}.}
  \bibinfo{year}{2019}\natexlab{}.
\newblock \showarticletitle{The Student Expectations of Learning Analytics
  Questionnaire}.
\newblock \bibinfo{journal}{\emph{Journal of Computer Assisted Learning}}
  \bibinfo{volume}{35}, \bibinfo{number}{5} (\bibinfo{year}{2019}),
  \bibinfo{pages}{633--666}.
\newblock
\urldef\tempurl%
\url{https://doi.org/10.1111/jcal.12366}
\showDOI{\tempurl}
\showeprint{https://onlinelibrary.wiley.com/doi/pdf/10.1111/jcal.12366}


\bibitem[Whitelock-Wainwright et~al\mbox{.}(2020)]%
        {SELAQ-multilang}
\bibfield{author}{\bibinfo{person}{Alexander Whitelock-Wainwright},
  \bibinfo{person}{Dragan Gašević}, \bibinfo{person}{Yi-Shan Tsai},
  \bibinfo{person}{Hendrik Drachsler}, \bibinfo{person}{Maren Scheffel},
  \bibinfo{person}{Pedro~J. Muñoz-Merino}, \bibinfo{person}{Kairit Tammets},
  {and} \bibinfo{person}{Carlos Delgado~Kloos}.}
  \bibinfo{year}{2020}\natexlab{}.
\newblock \showarticletitle{Assessing the validity of a learning analytics
  expectation instrument: A multinational study}.
\newblock \bibinfo{journal}{\emph{Journal of Computer Assisted Learning}}
  \bibinfo{volume}{36}, \bibinfo{number}{2} (\bibinfo{year}{2020}),
  \bibinfo{pages}{209--240}.
\newblock
\urldef\tempurl%
\url{https://doi.org/10.1111/jcal.12401}
\showDOI{\tempurl}
\showeprint{https://onlinelibrary.wiley.com/doi/pdf/10.1111/jcal.12401}


\bibitem[Winne(2017)]%
        {winne_learning_2017}
\bibfield{author}{\bibinfo{person}{Philip Winne}.}
  \bibinfo{year}{2017}\natexlab{}.
\newblock \showarticletitle{Learning {Analytics} for {Self}-{Regulated}
  {Learning}}.
\newblock In \bibinfo{booktitle}{\emph{The {Handbook} of {Learning}
  {Analytics}} (\bibinfo{edition}{1} ed.)},
  \bibfield{editor}{\bibinfo{person}{Charles Lang}, \bibinfo{person}{George
  Siemens}, \bibinfo{person}{Alyssa~Friend Wise}, {and} \bibinfo{person}{Dragan
  Gaševic}} (Eds.). \bibinfo{publisher}{Society for Learning Analytics
  Research (SoLAR)}, \bibinfo{address}{Alberta, Canada},
  \bibinfo{pages}{241--249}.
\newblock
\showISBNx{978-0-9952408-0-3}
\urldef\tempurl%
\url{http://solaresearch.org/hla-17/hla17-chapter1}
\showURL{%
\tempurl}


\bibitem[You(2016)]%
        {YOU201623}
\bibfield{author}{\bibinfo{person}{Ji~Won You}.}
  \bibinfo{year}{2016}\natexlab{}.
\newblock \showarticletitle{Identifying significant indicators using LMS data
  to predict course achievement in online learning}.
\newblock \bibinfo{journal}{\emph{The Internet and Higher Education}}
  \bibinfo{volume}{29} (\bibinfo{year}{2016}), \bibinfo{pages}{23--30}.
\newblock
\showISSN{1096-7516}
\urldef\tempurl%
\url{https://doi.org/10.1016/j.iheduc.2015.11.003}
\showDOI{\tempurl}


\end{thebibliography}

\end{document}